\documentclass[12pt]{article}
\usepackage{a4}
\usepackage{amsfonts}
\usepackage{amsmath,amssymb}

\usepackage[dvipdfmx]{graphicx}

\let\du=\du                     
        
\def\a{\alpha}
\def\b{\beta}

\def\k{\kappa}

\def\t{\tau}

\def\L{\Lambda}

\newcommand{\be}{\begin{equation}}
\newcommand{\ee}{\end{equation}}
\newcommand{\nbe}{\begin{equation*}}
\newcommand{\nee}{\end{equation*}}

\newcommand{\lb}{\label}

\begin{document}

\thispagestyle{empty}

{\hbox to\hsize{
\vbox{\noindent July 2016 \hfill IPMU16-0099 }}}

\noindent
\vskip2.0cm
\begin{center}

{\large\bf Randall-Sundrum Brane-World in Modified Gravity}
\vglue.3in

Hiroshi Nakada~${}^{a}$  and Sergei V. Ketov~${}^{a,b,c,d}$ 
\vglue.1in

${}^a$~Department of Physics, Tokyo Metropolitan University, \\
Minami-ohsawa 1-1, Hachioji-shi, Tokyo 192-0397, Japan \\
${}^b$~Kavli Institute for the Physics and Mathematics of the Universe (IPMU),
\\The University of Tokyo, Chiba 277-8568, Japan \\
${}^c$~Department of Physics, Faculty of Science, Chulalongkorn University,\\
Thanon Phayathai, Pathumwan, Bangkok 10330, Thailand\\
${}^d$~Institute of Physics and Technology, Tomsk Polytechnic University,\\
30 Lenin Ave., Tomsk 634050, Russian Federation \\
\vglue.1in
nakada-hiroshi1@ed.tmu.ac.jp, ketov@tmu.ac.jp
\end{center}

\vglue.3in

\begin{center}
{\Large\bf Abstract}
\end{center}
\vglue.1in
\noindent  We modify Randall-Sundrum model of brane-world (with two branes) by adding the scalar curvature squared term in five dimensions. We find that it does not destabilize Randall-Sundrum solution to the
hierarchy problem of the Standard Model in particle physics.

\newpage

\section{Introduction}

The basic idea of "brane-world" is an assumption that our "world" is confined to a "brane" in a higher-dimensional space-time. To the best of our knowledge, this idea was first proposed in 1983 by Rubakov and Shaposhnikov \cite{rush}  who studied a confinement of matter on a domain wall in a five-dimensional spacetime. In 1998, Arkani-Hamed, Dimopoulos and Dvali \cite{add} found that brane-world may be the alternative to (Kaluza-Klein) spacetime compactification in solving the hierarchy problem of particle physics.

Two elegant and simple brane-world models were proposed by Randall and Sundrum in 1999 \cite{rs}. The first
of them (dubbed RSI) has two branes embedded in a five-dimensional spacetime with a negative cosmological constant. One brane is identified with our visible Universe, whereas another brane is hidden, similarly to 
Horava-Witten model in string theory \cite{hwit}. The RSI  brane-world model generates a large (exponential) hierarchy between the electro-weak scale and the Planck scale by using the geometrical warp factor in a higher-dimensional spacetime metric, with a small extra dimension forming an orbifold $S^1/Z_2$ (see e.g., \cite{MKrev} for a review).

In brane-world scenarios, extra dimensions may be large and even infinite, while only gravity can be present
there. The brane-world idea  survives all experimental checks. However, there is no experimental evidence for the existence of extra dimensions either.

 The success of brane-world poses the question of its stability against possible modifications of Einstein gravity. Such modifications are inevitable because the Einstein gravity is known to be non-renormalizable, either in four
 or higher spacetime dimensions. In this paper we answer this question by modifying gravity in five spacetime
 dimensions via adding the simplest higher-order term proportional to the scalar curvature squared, in the
 context of the RSI brane-world model.
 
 We use the natural units, $c = \hbar = 1$, and the five-dimensional spacetime signature  
 $\eta_{\mu \nu} = {\rm diag} (-, +, + , + , +)$. Capital Latin indices refer to five spacetime dimensions, while lower case Greek indices refer to four dimensions of a brane. 
 
Our gravitational action in five dimensions reads
\begin{equation} \lb{grav}
 S_{\rm gr.}=\frac{1}{2\kappa^2}\int d^5x\sqrt{-g}(R+\alpha R^2- \Lambda)~~,
\end{equation}
where we have introduced  the gravitational coupling constant $\k$, the spacetime scalar curvature $R$,
the modified gravity coupling constant $\a$, and a cosmological constant $\Lambda$.
 
As in the RSI model \cite{rs}, we introduce a 'visible' brane (where SM fields are confined) located at $y=0$ with tension $\t_{\rm RS}$, and a 'hidden' brane located at $y=L$ and having the negative tension $-\t_{\rm RS}$, where the coordinate $y$ parametrizes extra dimension with the orbifold identification $-y\sim y$.
 Accordingly, our full action is given by
\begin{equation} \lb{fact}
 S=S_{\rm gr.}+S_{\rm vis.}+ S_{\rm hid.}~~,
\end{equation}
where 
\begin{eqnarray} \lb{bact}
 S_{\rm hid.}&=& -\t_{\rm RS} \int d^4xdy \sqrt{-g}\,\delta(y-L)~~,\\
 S_{\rm vis.}&=&\t_{\rm RS}\int d^4xdy\sqrt{-g}\,\delta(y)~~.
\end{eqnarray}

Our paper is organized as follows. In Sec.~2 we perform the duality transformation converting the modified gravity
into a scalar-tensor gravity, which greatly simplifies our treatment, because it allows us to get rid of the higher derivatives in the action (\ref{grav}). The scalar potential is studied in Sec.~3. In Sec.~4 we derive the 
brane-world equations for the warp factor coupled to extra dynamical scalar in five dimensions, and study their solutions.  Sec.~5 is our Conclusion.

\section{Duality transform to scalar-tensor gravity}

Let us replace  $R+\alpha R^2$ by $(1+2\alpha B)R-\alpha B^2$  in the gravitational action (\ref{grav}), by 
using a new scalar field $B$ in five dimensions, as
\begin{equation} \lb{eqgr}
S_{\rm gr.}=\frac{1}{2\kappa^2}\int d^5x\sqrt{-g}[(1+2\alpha B)R-\alpha B^2-\Lambda]~.
\end{equation}
The equation of motion of the field $B$ is algebraic and reads $B=R$. Hence, after substituting it back to the action (\ref{eqgr}) we get the equivalent action (\ref{grav}).

The action (\ref{eqgr}) can be brought to Einstein frame by a Weyl transformation in {\it five} dimensions 
({\it cf.} the similar transformation familiar in four dimensions, as was introduced in \cite{bick, whitt, maeda}),
\begin{equation} \lb{weyl}
 g_{AB}=\Omega^{-2}\tilde{g}_{ AB},\ \ \ \sqrt{-g}=\Omega^{-5}\sqrt{-\tilde{g}}~~,
\end{equation}
with a suitably chosen factor $\Omega$. By using the induced relation
\begin{equation}
 R=\Omega^2[\tilde{R}+8\tilde{\Box}f-12\tilde{g}^{AB}f_{,A}f_{,B}]~~,
\end{equation}
where the tildes refer to the transformed quantities, and the definitions
\begin{equation} \lb{defs}
 f=\ln \Omega\qquad {\rm and} \qquad f_{,A}=\frac{\partial_A\Omega}{\Omega}~~,
\end{equation}
we find the gravitational action as follows:
\begin{equation} \lb{trgr}
 S_{\rm gr.}=\frac{1}{2\kappa^2}\int d^5x\sqrt{-\tilde{g}}\Omega^{-5}[(1+2\alpha\phi)\Omega^2(\tilde{R}+8\tilde{\Box}f -12\tilde{g}^{AB}f_{,A}f_{,B})-\alpha B ^2-\Lambda]~.
\end{equation}
Hence, to get the transformed gravitational action in Einstein frame, we set 
\begin{equation} \lb{sete}
\Omega^{3}=e^{3f}=1+2\alpha B~.
\end{equation}
It yields
\begin{equation}\lb{fwb}
 f=\frac{1}{3}\ln(1+2\alpha B)
\end{equation}
and
\begin{equation} \lb{grein}
 S_{\rm gr.}=\frac{1}{2\kappa^2}\int d^5x\sqrt{-\tilde{g}}[\tilde{R}-12\tilde{g}^{AB}\partial_Af\partial_Bf
 -e^{-5f}(\alpha B^2+\Lambda)]~.
\end{equation}

A canonically normalised scalar kinetic term is obtained after rescaling
\begin{equation} \lb{resc}
\phi=2\sqrt{3}f/\kappa~,
\end{equation}
so that we find
\begin{equation} \lb{canb}
 B=\frac{ \exp\left(\frac{\sqrt{3}\kappa\phi}{2}\right)-1}{2\alpha}
\end{equation}
and the scalar potential
\begin{equation} \lb{pot}
V(\phi)=\frac{1}{8\alpha\kappa^2} \left[
\exp \left(\frac{\kappa\phi}{2\sqrt{3}}\right)
-2\exp \left(-\frac{\kappa\phi}{\sqrt{3}}\right)
+(1+4\alpha \Lambda)\exp\left( -\frac{5\kappa\phi}{2\sqrt{3}}\right)\right].
\end{equation}

As a result, the gravitational action takes the form
\begin{equation}\lb{aform}
 S_{\rm gr.}=\frac{1}{2\kappa^2}\int
  d^5x\sqrt{-\tilde{g}}\tilde{R}+\int d^5x\sqrt{-\tilde{g}}\left[
  -\frac{1}{2}\tilde{g}^{AB}\partial_A\phi\partial_B\phi-V(\phi)\right]~.
\end{equation}

Our main result of this Section is the scalar potential (\ref{pot}) induced by the modified
$(R+R^2)$ gravity in five spacetime dimensions.

\section{Scalar dynamics in RSI brane-world}

As is demonstrated in the previous Section, the net effect of adding the $R^2$ term to the gravitational
action amounts to the presence of the extra dynamical scalar field $\phi$ minimally coupled to gravity 
and having the scalar potential (\ref{pot}) in five dimensions.

Note that the five-dimensional cosmological constant $\Lambda$ enters both (\ref{pot}) and
(\ref{aform}) via the factor
\begin{equation} \lb{beta}
\beta=1+4\alpha\Lambda~,
\end{equation}
while the scalar potential is bounded from below provided that
\begin{equation} \lb{bbb}
 \beta > 0 ~.
\end{equation}
In addition, demanding the 5-dimensional cosmological constant to be negative, as is needed in the
RSI model, we get
\begin{equation} \lb{bbb}
 \beta \leq 1 ~.
\end{equation}

The profile of the scalar potential at $\b=1/2$ is given in Fig.~1.

\begin{figure}[htbp] 
\begin{center}
\includegraphics[clip, width=8cm]{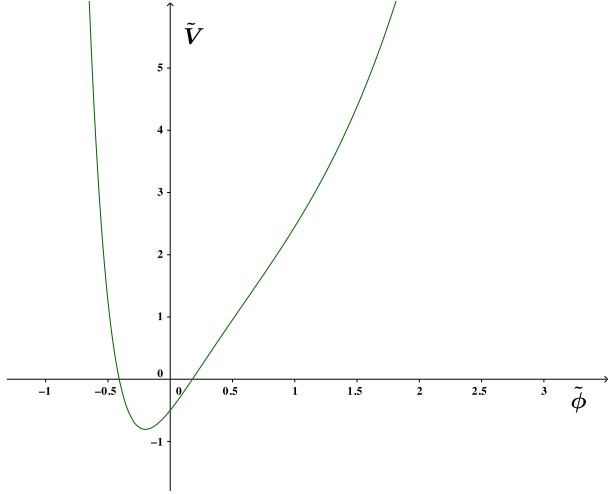}

\caption{The profile of the scalar potential $8\alpha\kappa^2 V(\phi)=\tilde{V}(\tilde{\phi})$ 
for $\tilde{\phi}=\frac{\kappa}{2\sqrt{3}}\phi$  and $\beta=0.5$. The value of the scalar potential 
at its minimum is given by $\tilde{V}(\tilde{\phi_0})\approx -0.81$. There are two solutions to $V=0$:
one at $\tilde{\phi}_{\rm left}\approx -0.4$ and another at $\tilde{\phi}_{\rm right}\approx 0.17$, on the
left and on the right of the AdS minimum, respectively.}
\label{fig:one}
\end{center}
\end{figure}

The minimum of the scalar potential (\ref{pot}) is achieved at 
\begin{equation} \lb{minp}
\tilde{\phi_0}=\frac{\kappa}{2\sqrt{3}}\phi_0 =\frac{1}{3}ln
\left(-2+ \sqrt{4+5\beta}\right)
\end{equation}
so that, in particular, $\tilde{\phi_0}=\phi_0=0$ for $\beta=1$.

The value of the scalar potential  at its minimum for any $0<\beta\leq 1$ is given by
\begin{equation}\lb{minv}
8\alpha\kappa^2 V(\phi_0)=\tilde{V}(\tilde{\phi}_0)=
\frac{-6\left(\sqrt{4+5\beta}-2-\beta\right)}{\left(-2+\sqrt{4+5\beta}\right)^{5/3}}~~.
\end{equation}
The $\tilde{V}_0(\tilde{\phi}_0)$ is an increasing negative function of $\beta$ for 
$0<\beta<1$, and it vanishes at $\beta=1$. 

Since $\b=1$ implies $\L=0$, we assume that  $0<\beta< 1$ in what follows. Then the value
of the scalar potential at its minimum, defined by (\ref{minv}), is always negative, which corresponds
to an AdS vacuum.

\section{Modified RSI model}

The Einstein equations for the action (\ref{aform}) in five dimensions read
\begin{equation}\lb{ein}
 \tilde{G}_{AB}=\tilde{R}_{ AB}-\frac{1}{2}\tilde{g}_{ AB}\tilde{R}=\kappa^2T_{AB}~,
\end{equation}
where we have introduced the total energy-momentum tensor 
\begin{equation} \lb{tem}
 T_{AB}=T^{\phi}_{AB}+T^{\rm vis.}_{AB}+T^{\rm hid.}_{AB}
\end{equation}
as a sum of three contributions,
\begin{eqnarray} 
 T^{\phi}_{AB}&=&
 \partial_A\phi\partial_B\phi+\tilde{g}_{ AB}\left(-\frac{1}{2}\tilde{g}^{MN}\partial_M\phi\partial_N\phi-V(\phi)\right)~,\\
T^{\rm hid.}_{AB}&=&e^{-\frac{5\kappa}{2\sqrt{3}}\phi}\tilde{g}_{AB}\t_{ RS}\delta(y-L)~,\label{vis}\\
T^{\rm vis.}_{AB}&=&-e^{-\frac{5\kappa}{2\sqrt{3}}\phi}\tilde{g}_{ AB}\t_{ RS}\delta(y)~.\label{hid}
\end{eqnarray}

The RS Ansatz for the 5-dimensional spacetime metric with the 4-dimensional Poincar\'e symmetry 
is given by \cite{rs}
\begin{equation}\lb{rsa}
 ds^2=e^{-2A}\eta_{\mu\nu}dx^{\mu}dx^{\nu}+dy^2~,
\end{equation}
where the warp factor $A(y)$ only depends upon the coordinate $y$ of the hidden (fifth) 
dimension,~\footnote{We denote by $x^{\mu}$ the remaining (four) space-time coordinates.}
and $\eta_{\mu\nu}$ is Minkowski metric in four other dimensions.

The corresponding (non-vanishing) Christoffel symbols, Ricci tensor and Ricci scalar are
\begin{eqnarray}
\Gamma^5_{\mu\nu}&=&A'e^{-2A}\eta_{\mu\nu}~,\\
 R_{\mu\nu}&=&\left(A''-4{A'}^2\right)e^{-2A}\eta_{\mu\nu}~,\\
R_{55}&=&4A''-4{A'}^2~,\\
R&=&8A''-20{A'}^2~,
\end{eqnarray}
respectively, where the primes denote the derivatives with respect to $y$.

Therefore, the Einstein equations take the form
\begin{eqnarray} \lb{einst}
G_{55}&=&6(A')^2=\kappa^2 T_{55}=\kappa^2 \left[\frac{1}{2}(\phi')^2-V(\phi)\right]~,\\
G_{\mu\nu}&=&-3\left(A''+2{A'}^2\right)e^{-2A}\eta_{\mu\nu}=\kappa^2 T_{\mu\nu}~,\\
T_{\mu\nu}&=&\tilde{g}_{\mu\nu}\left[-\frac{1}{2}\tilde{g}^{55}(\phi')^2-V(\phi)\right]\\
 &+&e^{-\frac{5\kappa}{2\sqrt{3}}\phi}e^{-2A}\eta_{\mu\nu}\t_{\rm RS}\delta(y-L)\\
 &-&e^{-\frac{5\kappa}{2\sqrt{3}}\phi}e^{-2A}\eta_{\mu\nu}\t_{\rm RS}\delta(y)~,
\end{eqnarray}
where we have assumed that the scalar field $\phi$ only depends upon $y$, because it is also
required by the four-dimensional Poincar\'e invariance (on the visible brane).

The equation of motion of the scalar field $\phi(y)$ coupled to gravity reads
\begin{equation} \lb{kge}
 \phi''+4A'\phi'+\frac{V(\phi)}{d\phi}=0~.
\end{equation}

As is already seen in Fig.~1, the scalar potential is very (exponentially) steep, whereas the right-hand-side
of Eq.~(\ref{einst}) must be positive, so that a slow roll is impossible. Therefore, the scalar field is quickly {\it stabilized} 
at its minimum (AdS vacuum) $\phi_0$ in five dimensions.

We numerically verified the scalar stabilization with the potential (\ref{pot}) by using Runge-Kutta method and the
Maxima software in application to a system of two coupled ordinary differential equations (32) and (\ref{kge}). When
choosing the parameters as $\beta=1/2$ (as in Fig.~1) and $\alpha=1/96$ (it fixes the scale of $V$ in Eq.~(\ref{kge})), the initial conditions  $\phi' =V=A'=0$ are consistent with (32). The corresponding numerical solutions to the functions 
$\tilde{\phi}(y)$ and  $A'(y)$  are given in Fig.~2.
We conclude that the functions  $\tilde{\phi}(y)$ and  $A'(y)$ quickly approach constant values indeed, independently
upon their initial conditions.
\vglue.2in

\begin{figure}[htbp] 
\begin{center}
\begin{tabular}{cc}
\begin{minipage}{0.5\hsize}
\begin{center}
\includegraphics[clip, width=7cm]{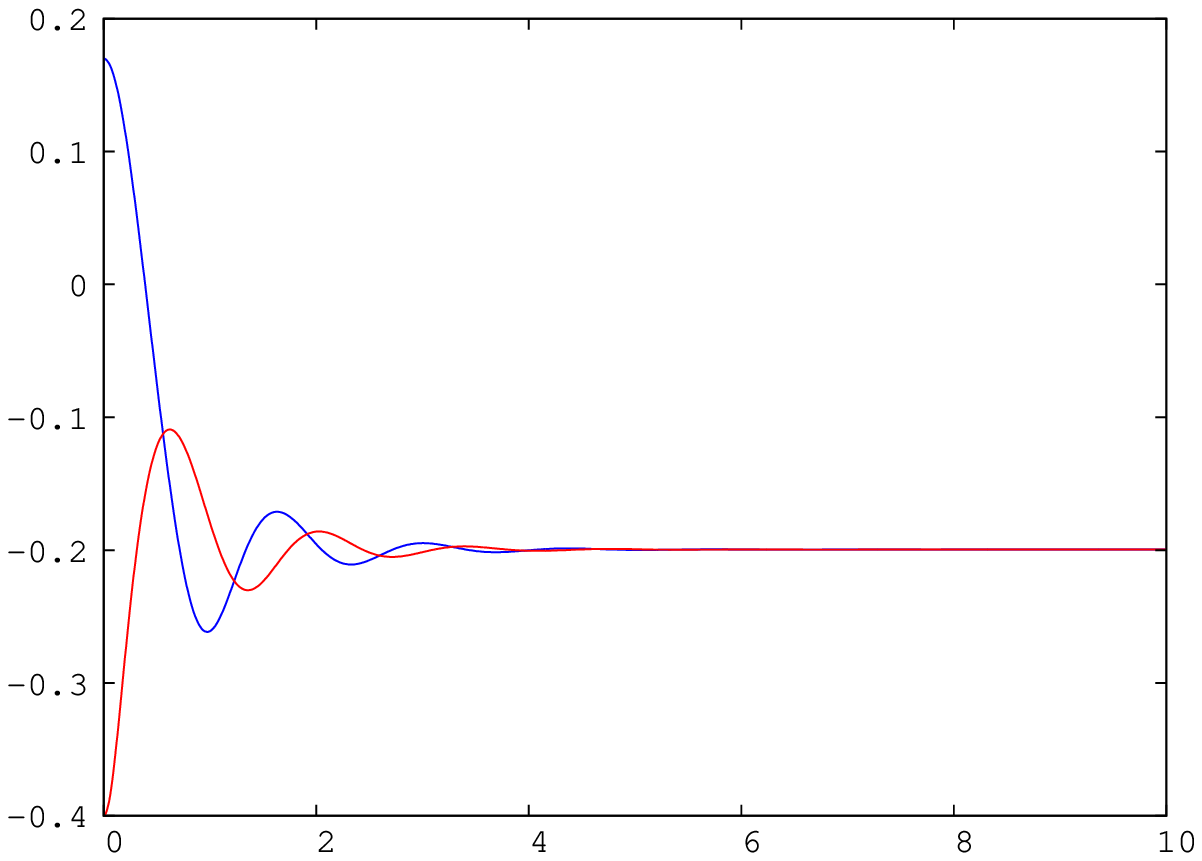}
\hspace{1.8cm}(a)
\end{center}
\end{minipage}
\begin{minipage}{0.5\hsize}
\begin{center}
\includegraphics[clip, width=7cm]{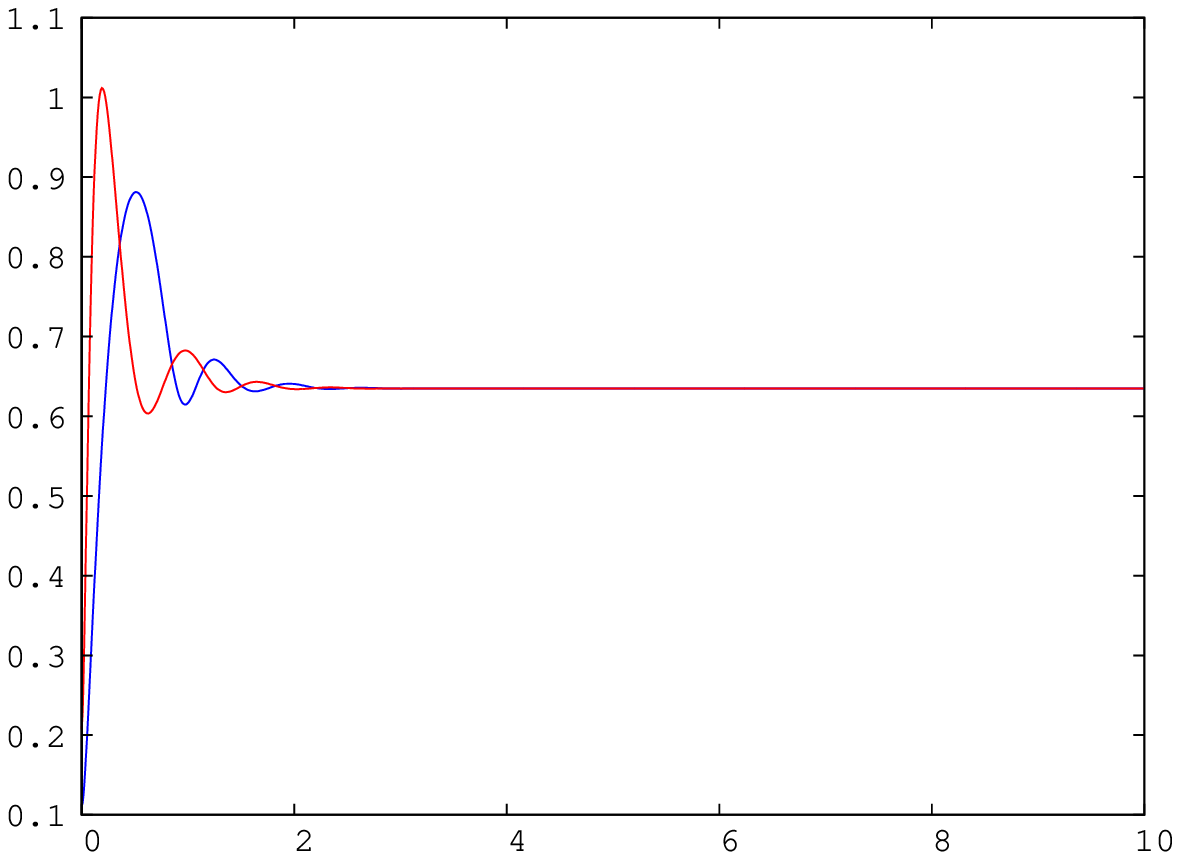}
\hspace{1.8cm}(b)
\end{center}
\end{minipage}
\end{tabular}
\caption{(a) the functions $\tilde{\phi}(y)$, and (b) the functions $A'(y)$, with the "left" (red) and "right" (blue) initial conditions, respectively.}
\end{center}
\label{fig:two}
\end{figure}

Once the scalar field is at its minimum (AdS vacuum) $\phi_0$ in five dimensions, the standard RSI scenario applies, 
being described by the solution \cite{rs}
\begin{equation} \label{rssol}
A(y)=k|y|~,\qquad -L\leq y \leq L~,\quad k>0~.
\end{equation}
In particular, the size $L$ of the 5th dimension is related to the Planck mass $M_{\rm Pl}$ in four dimensions
as
\begin{equation} \label{tofour}
\kappa^2 M_{\rm Pl}^2=\frac{1-e^{-2kL}}{2k}~.
\end{equation}
All that is consistent with the other Einstein equations provided that
\begin{equation} \lb{cons}
k^2 = - \k^2 V(\phi_0)\equiv - \k^2 V_{\rm min} \quad {\rm and} \quad \t_{\rm RS}=6k/\kappa^2~~.
\end{equation}
Then the negative cosmological constant in five dimensions is compensated on the visible brane by its tension, so
that we get Minkowski geometry there.

It is worth mentioning that the effective value of $\t_{\rm RS}$ gets modified in our
approach against the RSI one by the factor of $e^{-\frac{5\kappa}{2\sqrt{3}}\phi_0}$, because of Eqs.~(\ref{vis}) and
(\ref{hid}).

As in Ref.~\cite{rs}, the hierarchy between the electro-weak scale and the Planck scale 
$M_{\rm Pl}$ in four dimensions (on the visible brane) is achieved via the presence of the exponential factor 
$e^{-kL}$ induced by the extra dimension in the vacuum expectation value of Higgs field, so that we must require $kL\approx \ln 10^{16}\approx 35$. Then the exponential term in Eq.~(\ref{tofour}) becomes very small and can be ignored. The Newtonian limit of RSI model leads to the similar exponentially small corrections to Newton law of gravity, which are not in conflict with observations.

\section{Conclusion}

Given the phenomenological viability of Randall-Sundrum brane world, as the established and reasonable alternative to Kaluza-Klein compactification, it makes sense to analyse stability of the RS brane-world against possible modifications of gravity, as well as against quantum gravity corrections. In our investigation, we did a small step in this direction by proving stability of the RSI model against the simplest modification of the higher-dimensional gravity described by adding the scalar curvature squared term in five dimensions. 

An impact of the $R^2$-modified gravity on the RSI model can be simply described in terms of a single dynamical scalar field with the particular scalar potential (\ref{pot}). It is clear from our construction that this scalar has the gravitational origin
as spin-0 part of five-dimensional spacetime metric. We found that the value of the RSI parameter $k$ is determined
by dynamics of that scalar in the fifth dimension.

\section*{Acknowledgements}

SVK was supported by a Grant-in-Aid of the Japanese Society for Promotion of Science (JSPS) under No.~2640025200, a TMU President Grant of Tokyo Metropolitan University in Japan, the World Premier International Research Center Initiative (WPI Initiative), MEXT, Japan, the CUniverse research promotion project by Chulalongkorn University (grant reference CUAASC) in
Bangkok, Thailand, and the Competitiveness Enhancement Program of Tomsk Polytechnic University in Russia.

\end{document}